# Spin susceptibility and effective mass of two-dimensional electrons in $Mg_xZn_{1-x}O/ZnO$ heterostructures

A. Tsukazaki,[1] A. Ohtomo,[1] M. Kawasaki,[1,2,3] S. Akasaka,[4] H. Yuji,[4] K. Tamura,[4] K. Nakahara,[4] T. Tanabe,[4] A. Kamisawa,[4] T. Gokmen,[5] J. Shabani,[5] and M. Shayegan[5]

[1]*Institute for Materials Research, Tohoku University, Sendai 980-8577, Japan*
[2]*WPI Advanced Institute for Materials Research, Tohoku University, Sendai 980-8577, Japan*
[3]*CREST, Japan Science and Technology Agency, 102-0075, Japan*
[4]*Advanced Compound Semiconductors R&D Center, ROHM Co., Ltd., Kyoto 615-8585, Japan*
[5]*Department of Electrical Engineering, Princeton University, Princeton, New Jersey 08544*

We report measurements of the spin susceptibility and the electron effective mass for two-dimensional electrons confined at the interfaces of $Mg_xZn_{1-x}O/ZnO$ single heterostructures ($x$ = 0.05, 0.08, and 0.11), grown by molecular-beam epitaxy on (0001) ZnO substrates. By tuning the built-in polarization through control of the barrier composition, the electron density was systematically varied in the range of $5.6 \times 10^{11}$ - $1.6 \times 10^{12}$ cm$^{-2}$, corresponding to a range of $3.1 \leq r_s \leq 5.2$, where $r_s$ is the average electron spacing measured in units of the effective Bohr radius. We used the coincidence technique, where crossings of the spin-split Landau levels occur at critical tilt angles of magnetic field, to evaluate the spin susceptibility. In addition, we determined the effective mass from the temperature dependence of the Shubnikov-de Haas oscillations measured at the coincidence conditions. The susceptibility and the effective mass both gradually increase with decreasing electron density, reflecting the role of electron-electron interaction.

The quantum Hall effect (QHE) has been intensively investigated in two-dimensional electron systems (2DESs) in $Si/SiO_2$ and III-V compound semiconductors.[1-5] Recent technological advancements of ZnO epitaxy have enabled the growth of high-quality heterostructures exhibiting QHE.[6,7] Bulk ZnO has a direct band gap of 3.37 eV, an electron effective mass $m_b = 0.29m_0$,[8] and an effective Landé $g$ factor $g_b = 1.93$.[9] In an interacting system, $m_b$ and $g_b$ are renormalized to $m^*$ and $g^*$. So far, there is only one example of the effective mass measurement, reporting $m^* = 0.32 \pm 0.03\ m_0$ in $Mg_{0.2}Zn_{0.8}O/ZnO$.[6] For the observation of quantized magnetotransport, two criteria of $\omega_c\tau > 1$ and $\hbar\omega_c > k_BT$ must be fulfilled, where $\omega_c = eB/m^*$ is the cyclotron frequency, $e$ is the elementary charge, $B$ is the magnetic field, $\tau$ is the relaxation time, $\hbar$ is Planck's constant divided by $2\pi$, $k_B$ is Boltzmann's constant, and $T$ is the absolute temperature. The Landau levels are separated by cyclotron energy $E_C = \hbar eB_\perp/ m^*$, where $B_\perp$ is the perpendicular component of total magnetic field ($B_{tot}$). In addition to $m^*$, $g^*$ is also one of the most important parameters for 2D carriers as magnetotransport at high magnetic field is governed not only by the Landau levels but also by the Zeeman splitting energy, which increases with $B_{tot}$; $E_Z = g^*\mu_B B_{tot}$, where $\mu_B$ is the Bohr magneton. Note that in a Fermi liquid picture, the spin susceptibility ($\chi_s$) of the 2DES is in fact proportional to the product $g^*m^*$.

The interaction between electrons is an intriguing phenomenon in dilute 2DESs that results in, for example, an increase in $\chi_s$ and/or $m^*$. Indeed, in various 2D carrier systems such as $Si/SiO_2$,[10-16] GaAs/AlGaAs,[17-21] and AlAs,[22] a systematic enhancement of $\chi_s$ and/or $m^*$ as the density is lowered has been reported. Usually, the ratio of the Coulomb energy to the Fermi energy of the 2DES is used to describe the strength of electron-electron interaction with dimensionless parameter $r_s$. The value of $r_s$ is defined as $r_s = 1/\sqrt{\pi n} a_B^*$ and represents the average inter-electron spacing, measured in units of the effective Bohr radius; $n$ is the density of 2DES, $a_B^* = (\varepsilon / m_b) a_B$, $\varepsilon$ is the dielectric constant ($\varepsilon = 8.3$ for ZnO), and $a_B = 0.529$ Å is the hydrogen Bohr radius. Thus, the maximum range of $r_s$ attainable in particular semiconductor materials is the subject of interest; $r_s$ is relatively large in the present system owing to the small $a_B^*$ (18 Å).

Here we report on the measurements of $g^*m^*$ and $m^*$ for 2DESs confined at the $Mg_xZn_{1-x}O/ZnO$ heterointerfaces. These values are evaluated for five samples having 2DES density ranging from $5.6 \times 10^{11}$ to $1.6 \times 10^{12}$ cm$^{-2}$ ($3.1 \leq r_s \leq 5.2$). Our measurements demonstrate that $g^*m^*$ and $m^*$ increase with decreasing electron density, reflecting the strong electron-electron interaction in our system.

$Mg_xZn_{1-x}O/ZnO$ single heterostructures were pseudomorphically grown on Zn-polar ZnO single crystal substrates (Tokyo Denpa Co. Ltd.) by plasma-assisted molecular beam epitaxy.[7,23] At the $Mg_xZn_{1-x}O/ZnO$ interface, unintentionally doped electrons were accumulated by positive interfacial charges induced by the polarization mismatch between the constituents layers, which primarily depends on the composition of the heterostructure.[24-26] The parallel contribution of semiconducting bulk substrate to the conductance became negligibly small as the parasitic free electrons were frozen out at low temperature.[7] Consequently, the 2DES density could be controlled by tuning the Mg content in barrier layer without employing the modulation doping technique. In this study, we prepared five samples having different layered structures, *i.e.* Mg content in the barrier layer and thicknesses of the barrier and ZnO homoepitaxial layers, as listed in Table I. In the case of sample C, we employed a polymer Schottky contact,[27] which was electrically isolated from the interface channel, to reduce the 2DES density owing to surface depletion. Ohmic contacts were made by electron beam evaporation of Ti/Au. Using conventional photolithography and dry etching techniques, the samples were processed into a Hall-bar geometry (60 × 260 μm$^2$). Magnetotransport measurements were carried out by using ac lock-in technique with an excitation current of 50 nA in a $^3$He refrigerator with a base temperature of 0.3 K. The 2DES density evaluated from the low-field Hall effect and mobility ($\mu$) are also listed in Table I. A mobility range of 9000 to 20000 cm$^2$/Vs was achieved in our samples, corresponding to scattering times ($\tau$) of 1.5 to 3.4 ps.

TABLE I. The parameters for sample structures and transport properties at 0.3 K.

| Sample | Thickenss of homoepitaxial ZnO layer (nm) | Mg content ($x$) | Thickness of MgZnO layer (nm) | $n$ ($10^{12}$ cm$^{-2}$) | $\mu$ (cm$^2$V$^{-1}$s$^{-1}$) | $r_s$ |
|---|---|---|---|---|---|---|
| A | 100 | 0.05 | 100 | 0.87 | 19700 | 4.1 |
| B | 0 | 0.05 | 250 | 0.82 | 15000 | 4.3 |
| C | 100 | 0.05 | 300 | 0.56 | 20000 | 5.2 |
| D | 0 | 0.08 | 280 | 1.3 | 11000 | 3.4 |
| E | 100 | 0.11 | 190 | 1.6 | 9000 | 3.1 |

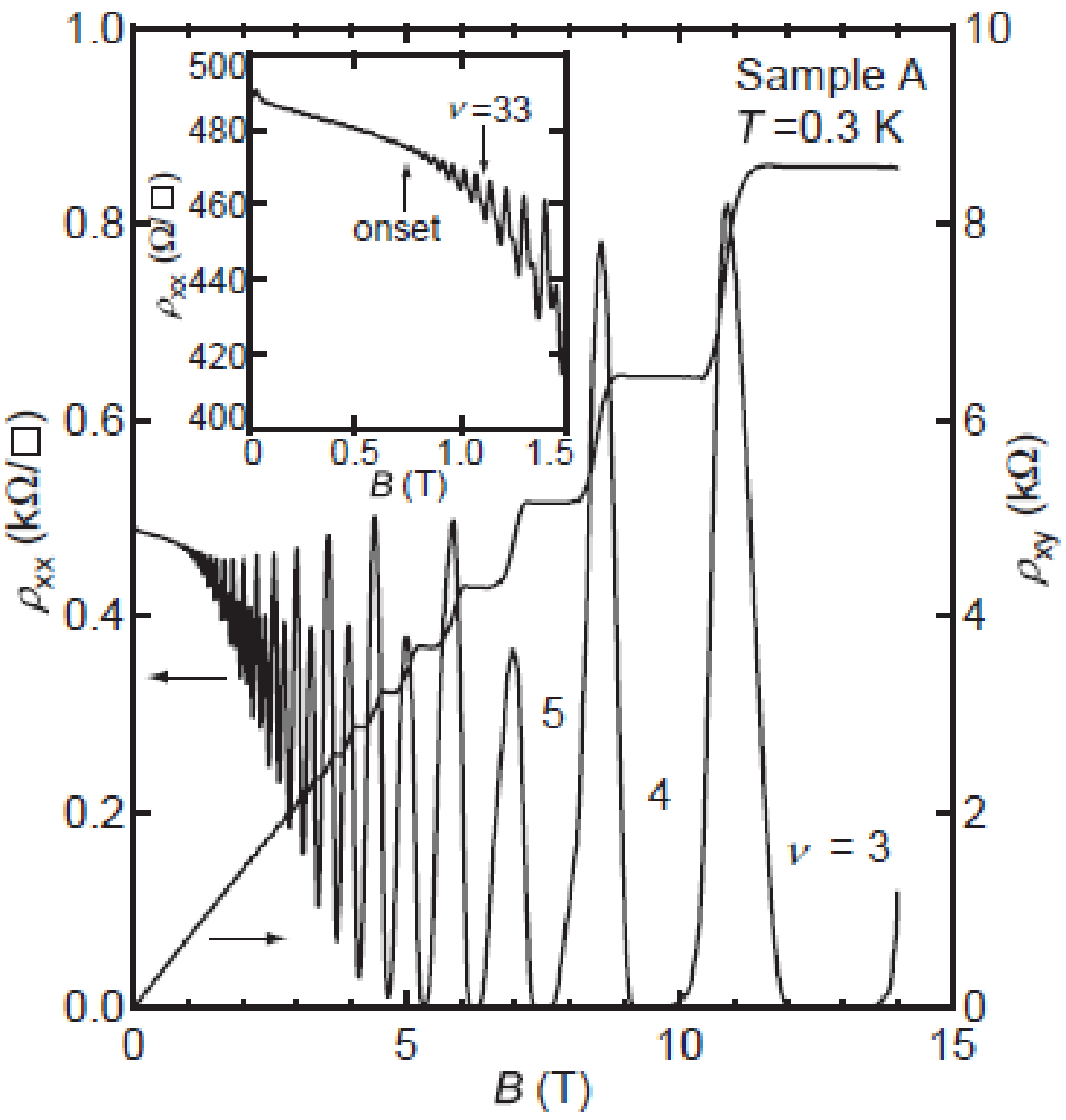

FIG. 1 Longitudinal resistivity $\rho_{xx}$ and Hall resistivity $\rho_{xy}$ vs $B$ measured for sample A at 0.3 K. Inset depicts $\rho_{xx}$ at low magnetic field. The oscillations start developing at an onset of 0.7 T with minima at odd Landau filling factors $\nu$.

Figure 1 shows typical longitudinal resistivity $\rho_{xx}$ and Hall resistivity $\rho_{xy}$ as a function of magnetic field applied perpendicular to the plane (sample A at 0.3 K). As shown in the inset, Shubnikov-de Haas oscillations start at an onset of ~0.7 T exhibiting $\rho_{xx}$ minima corresponding to odd Landau filling factor ($\nu$), and spin-split even-$\nu$ minima gradually develop at higher fields. The observation of stronger resistivity minima at odd-$\nu$ at the lowest fields is similar to the previous results obtained for a narrow AlAs quantum well,[28] and reflects the fact that in these 2DESs the cyclotron energy is close to the Zeeman energy (at zero tilt angle). Above 5T, $\rho_{xx}$ vanishes at integer $\nu$ and $\rho_{xy}$ exhibited Hall plateaus, reflecting the signatures of fully developed QHE states. The lowest $\nu$ attained in our samples were 3 and 2 for samples A and C, respectively. The electron density derived from the oscillation period agreed with $n_{\mathrm{Hall}}$ for all the samples within the accuracy of our measurements, indicating that parallel bulk conduction is negligible.

In order to evaluate $g^*m^*$, we used the coincidence technique, where $\rho_{xx}$ was recorded at 0.3 K under a magnetic field $B_{\mathrm{tot}}$ at various tilt angles $\theta$ as shown in Fig. 2(a). The perpendicular component $B_\perp$ is defined as $B_\perp = B_{\mathrm{tot}}\cos\theta$. The value of $\theta$ was calibrated by the simultaneous $\rho_{xy}$ measurements, since the total electron density does not change while tilting. Figure 2(b) explains the evolution of electron energy levels in a constant $B_\perp$ as $B_{\mathrm{tot}}$ and/or $\theta$ is increased. As the Zeeman splitting energy $E_Z$ increases with increasing $B_{\mathrm{tot}}$, crossing of energy levels are realized at the so-called coincidence angles. As a result, $E_Z/E_C$ changes as $E_Z/E_C = g^*m^*/2\cos\theta = i$, where $i$ takes on integer values at coincidence angles.[29] Figure 2(c) displays $\rho_{xx}$ vs $B_\perp$ at different tilt angles. We observe $\rho_{xx}$ minima for both

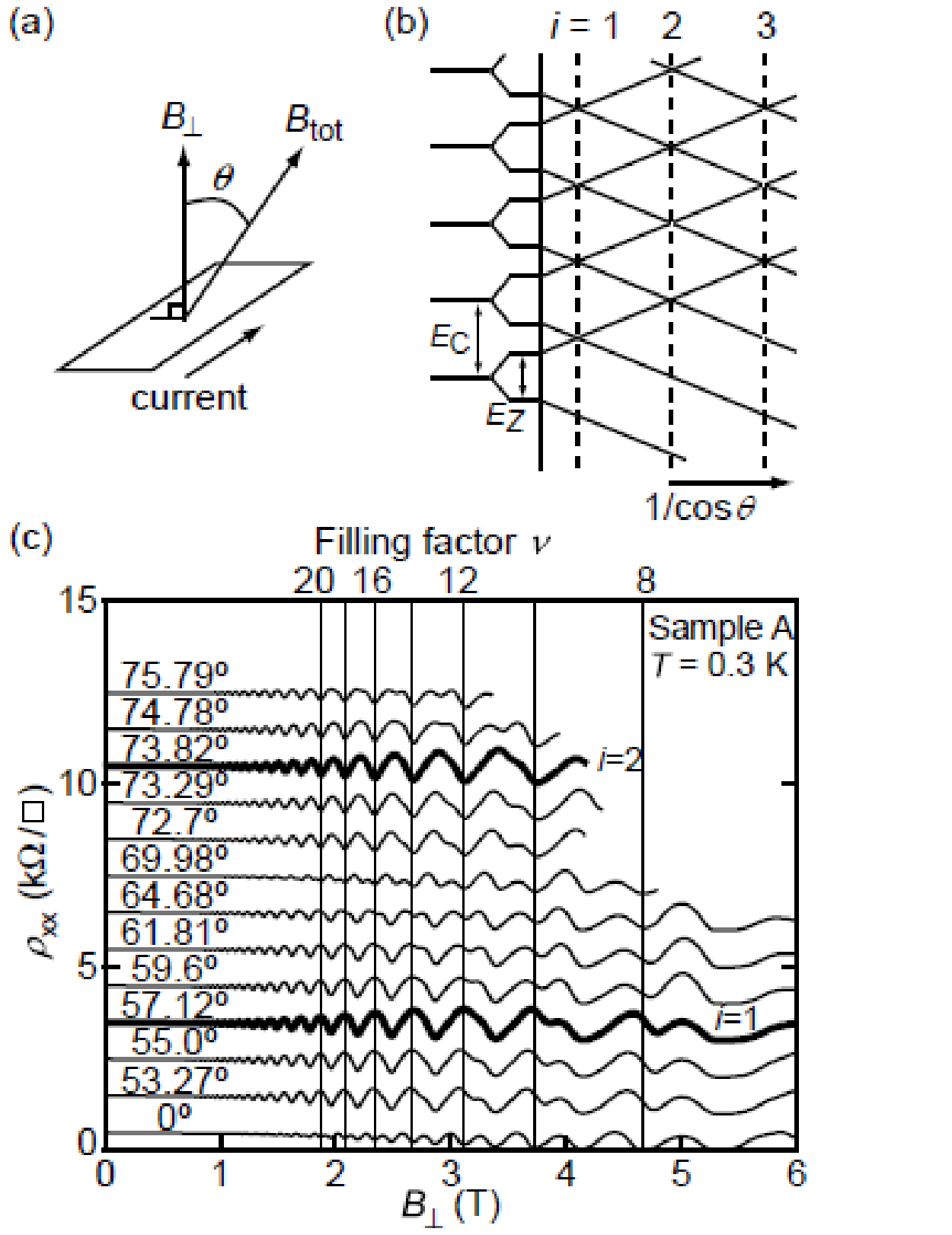

FIG. 2 (a) Measurement configuration in tilted magnetic field. (b) Schematic fan diagram showing spin-split Landau levels as a function of tilt angle $\theta$. (c) $\rho_{xx}$ vs $B_\perp$ measured for sample A at 0.3 K. Data taken at various tilt angles are shown and shifted vertically for clarity. Coincidences at $i$ = 1 and 2 are emphasized as bold traces.

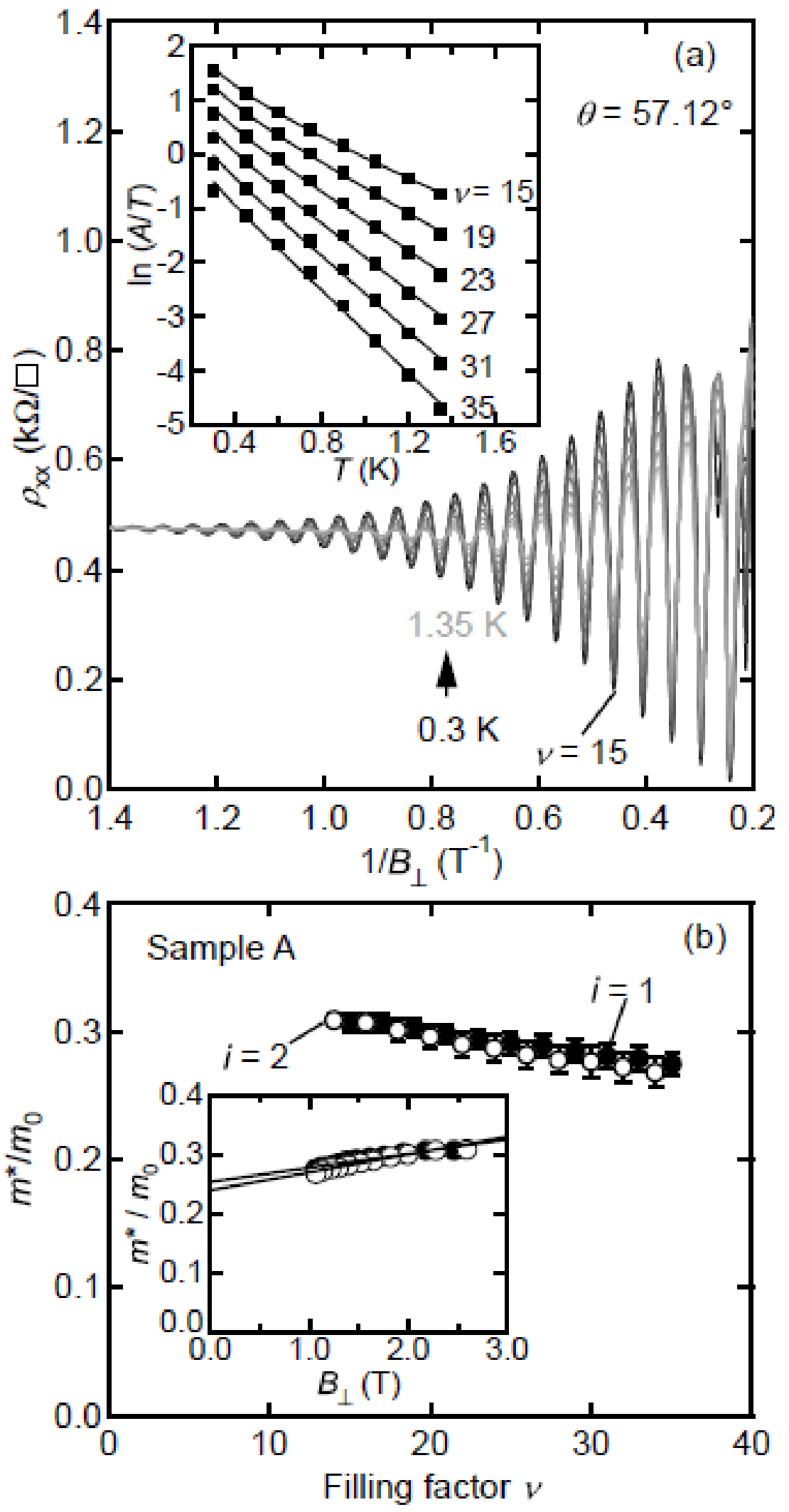

FIG. 3 (a) Temperature dependence of $\rho_{xx}$ vs $1/B_\perp$ traces recorded for sample A at the first coincidence angle ($\theta$ = 57.12°). Inset depicts the temperature dependence of the logarithmic amplitude of $\rho_{xx}$ minima at odd $\nu$ ranging from 15 to 35 (filled squares). Solid lines represent least-squares fits to the plots; the slopes of these lines give $m^*$. (b) The deduced $m^*/m_0$ at the first and second coincidence angles are plotted against $\nu$ as closed and open circles, respectively. The error bars are from standard deviations of the fits to ln($A/T$) vs $T$. Inset depicts $m^*/m_0$ plotted against perpendicular magnetic field $B_\perp$, where linear least-squares fits (solid lines) extrapolate to $m^*/m_0$ at $B_\perp = 0$.

odd and even $\nu$ at $\theta$ = 0° for $B_\perp$ > 1.5 T. As $\theta$ is increased, even-$\nu$ minima become weak, and those for $\nu \geq 12$ vanish at the first coincidence angle of 57.12° [bold trace in Fig. 2(c)], exhibiting crossing of the spin-split Landau levels. From this first coincidence angle, we deduce $g^*m^*$ = 1.1, almost twice larger than $g_bm_b$. The second coincidence is also identified at $\theta$ = 73.82°. It is worth mentioning that at lower Landau level fillings, e.g., at $\nu$ = 8, the resistance exhibits a minimum at the coincidence angle at the base temperature. This behavior reflects an anti-crossing of the Landau levels and the opening of an energy gap at the Fermi level. Similar anti-crossings have been reported in other 2DESs with relatively large $r_s$, and are attributed to electron-electron interaction.[30-33] We note that, in our case, as the temperature is raised, the resistance minimum quickly disappears and turns into a maximum; evidently, the anti-crossing gap in our samples is too small to be quantitatively measurable from data taken at $T$ > 0.3 K.

The value of $m^*$ was determined for the samples (A-D) from the Dingle analysis of the temperature dependence of the Shubnikov-de Haas oscillations. The oscillations in $\rho_{xx}$ were recorded in a temperature range from 0.3 to 1.35 K with an incremental change of 0.15 K. The tilt angle was fixed at the coincidence angles to ensure that the separation between adjacent energy levels is equal to the cyclotron energy. Figure 3(a) shows temperature dependence of $\rho_{xx}$ vs $1/B_\perp$ recorded for sample A at the first coincidence angle. Logarithmic amplitude of $\rho_{xx}$ minima plotted against $T$ were fitted using the Dingle formula, $\Delta\rho_{xx}/\rho_0 = 4\chi\exp(-\pi/\omega_c\tau_q)/\sinh\chi$, where $\chi = 2\pi^2k_BT/\hbar\omega_c$, and $\tau_q$ is the quantum life time. Assuming that $\tau_q$ has no temperature dependence in this range, we extracted $m^*$ at the first and second coincidence angles for odd and even $\nu$ states, respectively [Fig. 3(b)]. The error bars represent the standard deviation of the fits. The $\nu$ dependence of $m^*$ for both first and second coincidence data is quantitatively similar. As shown in inset of Fig. 3(b), the values of $m^*$ increase with increasing $B_\perp$ but their extrapolation to $B_\perp = 0$ gives nearly the same value. We do not know the origin of the dependence of $m^*$ on the magnetic field. We remark, however, that it is reminiscent of the field dependence of the effective mass for the light holes in the GaAs 2D hole systems.[34]

In Figure 4, we provide the density dependence of the values of $g^*m^*$ and $m^*$ normalized to the bulk values of $g_bm_b$ and $m_b$, respectively. The values were obtained for five samples from the same measurements and evaluation procedures as described above. As for $m^*/m_b$, we plotted the values at zero magnetic field, which were estimated from the linear extrapolations of $m^*/m_b$ vs $B_\perp$ [see Fig. 3(b) inset]. Both $g^*m^*$ and $m^*$ increase with decreasing 2DES density, especially $g^*m^*$ is enhanced more than twofold relative to $g_bm_b$. This result can be attributed to electron-electron interaction owing to the relatively large $r_s$, although the reason for smaller values of $m^*$ than the $m_b$ is yet unclear. The renormalization of $m^*$ and $g^*m^*$ in dilute 2D electron systems in semiconductors has been widely reported.[19,28,35,36] Typically, $m^*$ is enhanced compared to the bulk value at the lowest densities (largest $r_s$) and the enhancement increases with decreasing density. In the high density regime ($r_s \leq 3$), however, a suppression of $m^*$ to values slightly below $m_b$ has been reported both experimentally and theoretically (see ref. 19 and references therein). While the rise in $m^*$ that we observe in Fig. 4 as the density is decreased is consistent with previous results, a quantitative understanding of the $m^*$ suppression at $r_s \leq 5$, requires further work. Our enhanced values of $g^*m^* / g_b m_b$ shown in Fig. 4 are also in quantitative agreement with the results reported for other 2D systems.[28,35,36] The enhancement we observe is less than what is expected for an ideal 2D system with zero layer thickness; this difference is likely because of the finite thickness of the electron layer in our system.[35,36] We do not understand, however, why $g^*m^*$ in our highest density sample does not follow the trend of the other samples.

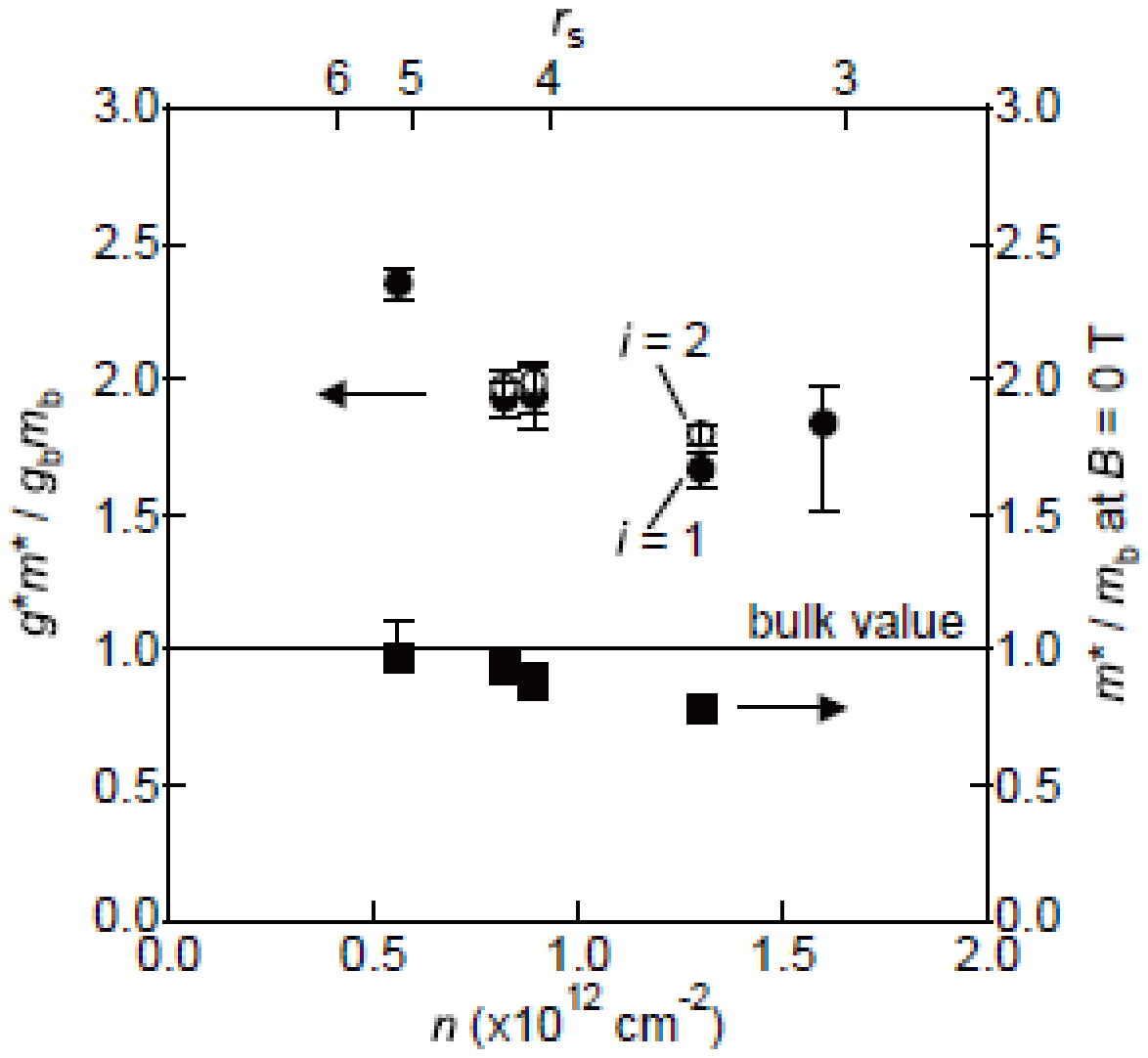

FIG. 4 $g^*m^*/g_bm_b$ and $m^*/m_b$ as a function of 2D electron density, $n$. Data are evaluated for samples A to E from measurements carried out at the first coincidence angles. In addition, $g^*m^*/g_bm_b$ for samples A, B, and D can be evaluated at the second coincidence angles. The one-sided error bars for $m^*$ represent the lowest value of $m^*$ measured from the low-field range and the extrapolated ($B$ = 0) value (see Fig. 3(b) inset). The normalized average inter-electron spacing $r_s$ is given in the top abscissa. The horizontal line marks bulk values for $g^*m^*$ and $m^*$.

In conclusion, we have evaluated $g^*m^*$ and $m^*$ for 2DESs confined in $Mg_xZn_{1-x}O$/ZnO heterostructures grown on Zn-polar ZnO substrates. We have observed crossing and anticrossing of spin-split Landau states at the coincidence conditions with a boundary near $\nu = 12$ for sample A with a density of $8.7 \times 10^{11}$ cm$^{-2}$. The value of $g^*m^*$ is about two times larger than bulk value and its dependence on the 2DES density is consistent with the role of the electron-electron interaction. Because the present results were obtained for samples having different structures and mobilities, possible effects of disorder or other difference between the samples cannot be excluded. External electric field control of 2DES density, which we plan for the future, will allow us to discuss electron-electron interaction, especially in the much higher $r_s$ region.

The authors thank to M. Nakano and K. Ueno for technical help for sample preparation. This research was partially supported by the Global COE Materials Integration Program, Tohoku University, the Japan Society for the Promotion of Science (JSPS), and the Asahi Glass Foundation. The work at Princeton University was supported by the NSF.